\documentclass[showpacs,preprint,aps,prx,preprint]{revtex4-1}
\usepackage{amsfonts}
\usepackage{amsmath}
\usepackage{amssymb}
\usepackage{graphicx}
\usepackage{txfonts}%
\usepackage{xcolor}%
\providecommand{\U}[1]{\protect\rule{.1in}{.1in}}

\begin{document}

\title{Classical-noise-free sensing based on quantum correlation measurement}
\today
\author{Ping Wang}
\author{Chong Chen}
\author{Renbao Liu}
\email{Corresponding author. Email: rbliu@cuhk.edu.hk}
\affiliation{Department of Physics and The Hong Kong Institute of Quantum Information Science and Technology, The Chinese University of Hong Kong, Shatin, New Territories, Hong Kong, China}

\begin{abstract}
Quantum sensing, using quantum properties of sensors, can enhance resolution, precision, and sensitivity of imaging, spectroscopy, and detection. An intriguing question is: Can the quantum nature (quantumness) of sensors and targets be exploited to enable schemes that are not possible for classical probes or classical targets? Here we show that measurement of the quantum correlations of a quantum target indeed allows for sensing schemes that have no classical counterparts. As a concrete example, in case where the second-order classical correlation of a quantum target could be totally concealed by non-stationary classical noise, the higher-order quantum correlations can single out a quantum target from the classical noise background, regardless of the spectrum, statistics, or intensity of the noise. Hence a classical-noise-free sensing scheme is proposed. This finding suggests that the quantumness of sensors and targets is still to be explored to realize the full potential of quantum sensing. New opportunities include sensitivity beyond classical approaches, non-classical correlations as a new approach to quantum many-body physics, loophole-free tests of the quantum foundation, et cetera.
\end{abstract}


\maketitle


\section{Introduction}
Quantum sensing or quantum metrology~\cite{Feynmanlecture,DegenRMP2017,PezzeRMP2018}
uses quantum properties (quantumness), such
as quantum coherence and quantum entanglement, of single or a few qubits to enhance detection
and measurement. Various quantum sensing schemes~\cite{GiovannettiScience2004,CavesPRD1981,BudkerNatPhys2007,JaklevicPRA1965,AbbottPRL2016,FagalyROSI2006,TaminiauPRL2012,KolkowitzPRL2012,ZhaoNatNano2012,PannetierScience2004,LaraouiNatCommun2013,SimonScience2017,BossScience2017,PfenderNC2019,DegenNature2019,TaminiauNature2019,DenkScience1990,MaletinskyNatNano2012,RugarNature2004,MochalinNNano2012,Steinert2013,AjoyPRA2012}
have been proposed and demonstrated useful to improve the detection
sensitivity~\cite{AbbottPRL2016,PannetierScience2004,FagalyROSI2006,ZhaoNNANO2011,TaminiauPRL2012,KolkowitzPRL2012,ZhaoNatNano2012},
 spectral resolution~\cite{LaraouiNatCommun2013,SimonScience2017,BossScience2017,PfenderNC2019,DegenNature2019,TaminiauNature2019},
and/or imaging resolution of metrological techniques~\cite{DenkScience1990,RugarNature2004,MochalinNNano2012,Steinert2013,MaletinskyNatNano2012,AjoyPRA2012},
including optical microscopy~\cite{DenkScience1990}, force microscopy~\cite{RugarNature2004},
bio-sensing~\cite{MochalinNNano2012}, magnetic resonance spectroscopy
and imaging~\cite{Steinert2013,MaletinskyNatNano2012}, navigation~\cite{AjoyPRA2012},
etc. An interesting question is: Can the quantumness of sensors and
targets be exploited to enable quantum sensing schemes that have no
classical counterparts? Below we discuss the rationales for a positive
answer.

First, we consider the $\emph{quantumness of sensors}$. We argue that it can be employed
to detect certain correlations of a quantum system (the target) that
are inaccessible to a classical probe. Information that can be obtained
from a target is all included in the response of the target to the
``force'' exerted from the sensor or probe. In general, the force
from a quantum sensor is a quantum operator (denoted as $\hat{S}$),
and that from a classical probe is a classical quantity $S$, coupled
to a \textquotedblleft displacement\textquotedblright{} operator $\hat{B}$
of the target, described by an interaction Hamiltonian $\hat{V}=-\hat{S}\hat{B}$
for a quantum sensor and $\hat{V}=-S\hat{B}$ for a classical probe.
The response of the target is described by the evolution of a density
operator $\hat{\rho}$ under the Liouville equation $\frac{d}{dt}\hat{\rho}=-i\left[\hat{V},\hat{\rho}\right]$,
where $\left[\hat{A},\hat{B}\right]\equiv\hat{A}\hat{B}-\hat{B}\hat{A}$
is the commutator. The commutator vanishes if either of the two operators
is a classical quantity. On the contrary, the anti-commutator $\left\{ \hat{A},\hat{B}\right\} \equiv\hat{A}\hat{B}+\hat{B}\hat{A}$
would not vanish but reduce to the usual product if either of the
quantities is a classical number. In quantum sensing, the response
is governed by $\left[\hat{S}\hat{B},\hat{\rho}\right]=\frac{1}{2}\left\{ \hat{S},\left[\hat{B},\hat{\rho}\right]\right\} +\frac{1}{2}\left[\hat{S},\left\{ \hat{B},\hat{\rho}\right\} \right]$,
which involves both the commutator and the anti-commutator between
the displacement operator and the target state operator; in classical
probe, the response of the target is governed by $S\left[\hat{B},\hat{\rho}\right]$,
which contains only the commutator. Therefore, in all conventional
metrology that uses classical probes like optical fields, scanning
tips, coils, etc., the only accessible information about a quantum
target is the correlations that contain only the commutators like
${\rm Tr}\hat{B}\left[\hat{B},\hat{\rho}\right]$, ${\rm Tr}\hat{B}\left[\hat{B},\left[\hat{B},\hat{\rho}\right]\right]$
, ${\rm Tr}\hat{B}\left[\hat{B},\left[\hat{B},\left[\hat{B},\hat{\rho}\right]\right]\right]$
, etc. (corresponding to linear, second-order, third-order susceptibilities,
etc.), where \textquotedblleft Tr\textquotedblright{} denotes the
trace of an operator. Quantum sensing, on the contrary, can extract
correlations that interweave commutators and anti-commutators such
as ${\rm Tr}\hat{B}\left[\hat{B},\left\{ \hat{B},\left[\hat{B},\hat{\rho}\right]\right\} \right]$,
which are classically inaccessible.

Then we consider the {\it quantumness of targets}. The information
about a classical variable $B(t)$ is in general correlations like
$\left\langle B_{0}B_{1}\cdots B_{n}\right\rangle $, where $\langle\cdots\rangle$
denotes averaging over ensemble of measurements and $B_{n}\equiv B(t_{n})$;
for a quantum target, the correlations in general contain a mixture
of commutators and ani-commutators like ${\rm Tr}\hat{B}_{3}\left[\hat{B}_{2},\left\{ \hat{B}_{1},\left[\hat{B}_{0},\hat{\rho}\right]\right\} \right]$.
While the terms containing only anti-commutators ${\rm Tr}\hat{B}_{n}\left\{ \cdots\hat{B}_{2},\left\{ \hat{B}_{1},\left\{ \hat{B}_{0},\hat{\rho}\right\} \right\} \cdots\right\} $
would reduce to classical correlations $\left\langle B_{0}B_{1}\cdots B_{n}\right\rangle $
when the target approaches to the classical limit, the terms that
contain commutators have no classical counterpart, which we shall
call $\emph{quantum correlations}$.

Both the quantumness of a sensor and the quantum correlations of a
target are useful resources for enabling quantum sensing schemes that have no
classical counterpart. It is conceivable that quantum sensors'
capability of extracting classically inaccessible correlations may
provide new approaches to quantum many-body physics. Another potential application
of measuring the quantum correlations is loophole-free test of the
quantum foundation using statistics of data that has no classical
explanations (e.g., statistics that violates bounds similar but not
limited to the Leggett-Garg inequality~\cite{LeggettGargPRL1985}).
In this paper, we demonstrate a non-trivial application of the quantumness
of targets, namely, a classical-noise-free sensing scheme, utilizing
the fact that classical noises, regardless of their specific properties,
do not contribute to the quantum correlations at all. We show that
higher order quantum correlations can single out a quantum target
from classical background noises. The quantum correlations can be
extracted, e.g., by sequential weak measurement~\cite{WangPRL2019}.

The scheme we present is sensing of a quantum object in the presence
of classical noise. Under realistic conditions, the ``displacement''
of the target coupled to the sensor is always superimposed with environmental
noise. Various techniques can be adopted to filter out the noise and to single out the contribution
of the target~\cite{ZhaoNNANO2011,TaminiauPRL2012,KolkowitzPRL2012,ZhaoNatNano2012}.
In particular, the dynamical
decoupling~\cite{Mehring1983,ViolaPRA1998,ZanardiPLA1999} control
on the sensor with designed timing can filter out slow noises and
pick up the target signals that have certain temporal or spectral
features~\cite{KuboJPSJ1954a,AndersonJPSJ1954,CywinskiPRB2008}. Not
surprisingly, these schemes depend on the specific properties of the
noises. For example, the dynamical decoupling schemes require
the noise to be slow (color noise with a hard spectral cutoff). Here
we propose that by measuring the high-order quantum correlations that
are absent in classical targets, one can extract the signals of a
quantum target excluding contributions from any classical noises,
regardless their intensity (weak or strong), statistics (Gaussian,
telegraph, or else), spectra (slow, fast, or even white), etc.. Utilizing
the full quantumness of targets is a new strategy to combat the noise
effects in quantum sensing.

\section{An illustrating model}

\subsection{Correlation sensing by a spin-1/2 sensor}

Without loss of generality, we consider a sensor spin-1/2 coupled
to a quantum target through a so-called pure-dephasing interaction
\[
\hat{V}=-\hat{S}_{z}\hat{B}(t),
\]
 where $\hat{S}_{z}$ is the sensor spin operator along the $z$ axis
and the field $\hat{B}(t)=\hat{B}_{{\rm Q}}(t)+B_{{\rm C}}(t)$, with the quantum target and the classical noise
indicated by the subscripts ``Q'' and ``C'', respectively. The target is assumed to be initially in
a state described by a density operator $\hat{\rho}_{{\rm Q}}$ and
the classical noise has a probability distribution $\rho_{{\rm C}}$
(as a functional of the noise $B_{{\rm C}}(t)$).
Without intention to unify the diversified terminology in literature, here we define {\em classical correlations} as contain only anti-commutators,
such as $C^{++}(t_{m},t_{n})\equiv\left\langle {\rm Tr}\hat{\mathcal{B}}^{+}(t_{m})\hat{\mathcal{B}}^{+}(t_{n})\hat{\rho}_{{\rm Q}}\right\rangle $
(with the time order $t_{m}\ge t_{n}$) and $C^{+++}(t_{k},t_{m},t_{n})\equiv\left\langle {\rm Tr}\hat{\mathcal{B}}^{+}(t_{k})\hat{\mathcal{B}}^{+}(t_{m})\hat{\mathcal{B}}^{+}(t_{n})\hat{\rho}_{{\rm Q}}\right\rangle $
(with the time order $t_{k}\ge t_{m}\ge t_{n}$), where $\hat{\mathcal{A}}^{+}\hat{B}\equiv\left\{ \hat{A},\hat{B}\right\} /2$
(essentially the anti-commutator), and $\langle\cdots\rangle$ denotes
averaging over all realizations of the classical noise. The classical
correlations have contributions from both the target and the classical
noise background. For example, the second order correlation $C^{++}(t_{m},t_{n})=C_{{\rm Q}}^{++}(t_{m},t_{n})+C_{{\rm C}}^{++}(t_{m},t_{n})$
, with $C_{{\rm Q}}^{++}(t_{m},t_{n})\equiv{\rm Tr}\hat{\mathcal{B}}_{{\rm Q}}^{+}(t_{m})\hat{\mathcal{B}}_{{\rm Q}}^{+}(t_{n})\hat{\rho}_{{\rm Q}}$
and $C^{++}(t_{m},t_{n})\equiv\left\langle B_{{\rm C}}^{+}(t_{m})B_{{\rm C}}^{+}(t_{n})\right\rangle $.
We define {\em quantum correlations} as contain at least one commutator, such as
$C^{+-}(t_{m},t_{n})\equiv\left\langle {\rm Tr}\hat{\mathcal{B}}^{+}(t_{m})\hat{\mathcal{B}}^{-}(t_{n})\hat{\rho}_{{\rm Q}}\right\rangle $
(for $t_{m}\ge t_{n}$) and $C^{+-+}(t_{k},t_{m},t_{n})\equiv\left\langle {\rm Tr}\hat{\mathcal{B}}^{+}(t_{k})\hat{\mathcal{B}}^{-}(t_{m})\hat{\mathcal{B}}^{+}(t_{n})\hat{\rho}_{{\rm Q}}\right\rangle $
(for $t_{k}\ge t_{m}\ge t_{n}$), where $\hat{\mathcal{A}}^{-}\hat{B}\equiv-i\left[\hat{A},\hat{B}\right]/2$
(essentially the commutator). Note that if ether $\hat{A}$ or $\hat{B}$
is a classical quantity, $\hat{\mathcal{A}}^+\hat{B}=\hat{A}\hat{B}=\hat{B}\hat{A}$
and $\hat{\mathcal{A}}^{-}\hat{B}=0$. Therefore, the classical noise
does not contribute to the quantum correlations.

In principle, one can measure any quantum correlations of the
target to exclude the effects of the classical noise. However,
some constraints are worth mentioning. First, some quantum
correlations may vanish under realistic conditions and/or for the
specific sensor-target coupling. For example, the quantum correlations
$C^{+-}$ and $C^{++-}$ (or any term with a commutator $\hat{\mathcal{B}}^{-}\hat{\rho}$ at the earliest time)
vanish or are extremely small when the target (such as a nuclear spin)
has frequencies much lower than the temperature and therefore $\hat{\rho}_{{\rm Q}}\propto1$
and $ \hat{\mathcal{B}}^{-}\hat{\rho}=0$. For the specific sensing model
we will consider later in this paper (a target spin $\hat{\mathbf I}$ under a field $B_z$ that is perpendicular to the coupling with a sensor spin, $\hat{I}_x\hat{S}_z$),
 the third order quantum correlation $C^{+-+}$ vanishes.
 Furthermore, the quantum correlation
used for classical-noise-free sensing should be chosen to
be an $\emph{irreducible}$ one. For example, the fourth-order quantum
correlation $C^{+-++}$ may be factorized
as $C^{+-++}(t_j,t_{k},t_m,t_n)=C^{+-}(t_j,t_{k})C^{++}(t_m,t_n)+\tilde{C}^{+-++}(t_j,t_{k},t_m,t_n)$
with $\tilde{C}$ denoting the irreducible part. In such a case,
the fourth-order correlation would be dominated by the second-order
terms and one should choose to measure $C^{+-}$ (if it is not zero). To illustrate
a generally applicable scheme of classical-noise-free sensing, we
will use the fourth-order quantum correlation
\[
C^{+--+}(t_j,t_{k},t_m,t_n)\equiv\left\langle {\rm Tr}\hat{\mathcal{B}}^{+}(t_j)\hat{\mathcal{B}}^{-}(t_{k})\hat{\mathcal{B}}^{-}(t_m)\hat{\mathcal{B}}^{+}(t_n)\hat{\rho}_{{\rm Q}}\right\rangle ,
\]
 which is irreducible for target systems with $\hat{\rho}_{{\rm Q}}\propto1$
(i.e., under high temperature).

\begin{figure}
\includegraphics[width=0.9\columnwidth]{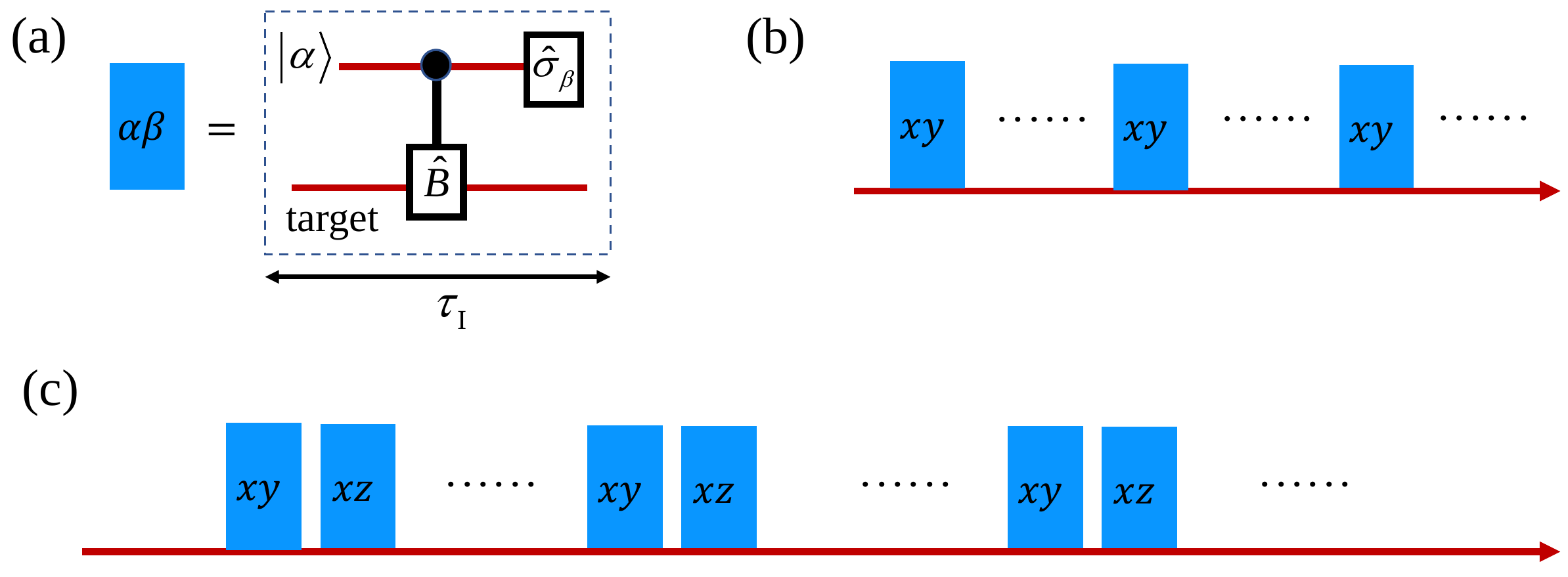}
\caption{Extraction of target correlations via sequential weak measurement.
(a) A shot of weak measurement (labeled as $\alpha\beta$, with, e.g.,
$\alpha=x$ and $\beta=y$), realized by first preparing the sensor
in the state $|\alpha\rangle$, then coupling the sensor and the target
through interaction $\hat{V}(t)=-\hat{S}_{z}\hat{B}(t)$ for a short
period of time $\tau_{{\rm I}}$, and finally measuring the sensor
operator $\hat{\sigma}_{\beta}$. (b) A sequence composed of repeated
measurement cycles $xy$. (c) A sequence composed of repeated measurement
cycles $xy$ and $xz$. (b) and (c) can be used for, e.g., extracting the second-order correlation $C^{++}$ and the fourth-order one $C^{+--+}$,
respectively.\label{fig_measurement}}
\end{figure}

The correlations of the target can be extracted from the correlations
of sequential weak measurement~\cite{WangPRL2019}. A shot of weak
measurement on the target can be realized by first preparing the sensor
spin-1/2 in the state $|\alpha\rangle$, then coupling the sensor
to the target through $\hat{V}=-\hat{S}_{z}\hat{B}(t)$ for a short
period of time $\tau_{\rm I}$, and finally measuring the sensor operator
$\hat{\sigma}_{\beta}$. Fig.~\ref{fig_measurement}(a) illustrates the process.
When $\left|\hat{B}\tau_{\rm I}\right|\ll1$, the entanglement between the
sensor and the target is small, so the projection measurement on the
sensor constitutes a weak measurement of the target. A sequence of
weak measurements consists of many ($M\gg1$) repeated cycles {[}examples
shown in Fig.~\ref{fig_measurement}(b)
and Fig.~\ref{fig_measurement}(c){]}.
The output of the
$\alpha\beta$-measurement shot in the $m$-th cycle is denoted as
$s_{m}^{\alpha\beta}$ (which takes value $+1$ or $-1$). The correlations
of the outputs are obtained as
\[
G_{\beta_{j}\cdots\beta_{m}\beta_{n}}^{\alpha_{j}\cdots\alpha_{m}\alpha_{n}}\left(t_{j},\ldots,t_m,t_n\right)=\frac{1}{M}\sum_{i}s_{j+i}^{\alpha_{j}\beta_{j}}\cdots s_{m+i}^{\alpha_{m}\beta_{m}}s_{n+i}^{\alpha_{n}\beta_{n}},
\]
where $t_{m}=m\tau_{\rm I}$.
Choosing properly the initial state and the measurement
basis in each cycle, arbitrary correlations of the target $C$ can
be extracted from the output correlation $G$~\cite{WangPRL2019}.
For example, in the second order
\[
G_{yy}^{xx}\left(t_m,t_n\right)=\tau_{\rm I}^{2}C^{++}\left(t_m,t_n\right)+O\left(\tau_{\rm I}^{4}\right),
\]
 which can be obtained using the measurement sequence shown in
Fig.~\ref{fig_measurement}(a).  Similarly, the fourth-order correlation
\[
G_{yzzy}^{xxxx}\left(t_j,t_{k},t_m,t_n\right)=\tau_{\rm I}^{4}C^{+--+}\left(t_j,t_{k},t_m,t_n\right)+O\left(\tau_{\rm I}^{6}\right),
\]
which can be obtained using the sequence in Fig.~\ref{fig_measurement}(c).
Since when the target is classical, i.e., $\hat{B}(t)$
is a classical quantity, the quantum correlation $C^{+--+}=0$,
the measurement output correlation $G_{yzzy}^{xxxx}=0$
in the leading order of the interaction time $\tau_{{\rm I}}$. This
result has a clear physical meaning. In the pure dephasing model,
if the sensor spin is prepared initially in the $x$ direction, it
would always be precessing in the $x$-$y$ plane and therefore the
measurement along the $z$ axis would always have 50:50 probability
ratio for the outputs $+1$ and $-1$, independent of the measurement
at other times. Thus, it seems that the output correlation $G_{yzzy}^{xxxx}(t_{j},t_{k},t_{m},t_{n})$
would always be zero. This is indeed the case if the field $\hat{B}(t)$
is classical. However, when the target is quantum, the measurements
along the $z$ axis at $t_{m}$ and $t_{k}$ would (weakly) polarize
the target through quantum back-action (i.e., state collapse due to
quantum measurement), with (slightly) higher probability in certain
states depending on the measurement outputs. The polarized target
state would affect the precession of the sensor spin afterwards and
hence the output of the measurement along the $y$ axis at time $t_{j}$,
inducing a non-vanishing correlation(see Appendix~\ref{Appendix_G4}
for more details). Actually, from the analysis above, the conclusion that the output
correlation $G_{yzzy}^{xxxx}$ vanishes for classical targets is valid
not only for the short interaction time limit.

\subsection{Sensing a spin-1/2 target}

As a concrete example, we consider the detection of a target spin-1/2
via a qubit sensor {[}Fig.\ref{fig_c2c4}. (a){]}. This scenario is
frequently encountered in sensing single nuclear spins~\cite{ZhaoNNANO2011,TaminiauPRL2012,KolkowitzPRL2012,ZhaoNatNano2012, LovchinskyScience2016,AslamScience2017,SushkovPRL2014}.
In diamond quantum sensing, for example, the sensor can be a shallow
nitrogen vacancy centre and the target can be a proton spin on the
diamond surface~\cite{StaudacherScience2013,MaminScience2013}. We
assume the target spin has an intrinsic energy splitting $\omega_{0}$
along the $z$ axis and its $x$ component couples to the target.
Thus, the target-sensor interaction takes the form
\[
\hat{V}=-\hat{S}_{z}\hat{B}(t)=-\hat{S}_{z}\left[2a\hat{I}_{x}(t)+B_{\mathrm{C}}(t)\right],
\]
where $\hat{I}_{x}(t)=e^{i\omega_{0}\hat{I}_{z}t}\hat{I}_{x}e^{-i\omega_{0}\hat{I}_{z}t}$
is the target spin operator in its interaction picture. The target-sensor
coupling coefficient $a$ can be tuned by dynamical coupling and decoupling~\cite{ViolaPRA1998,ZanardiPLA1999,Mehring1983}.
The classical noise $B_{{\rm C}}(t)$ acting on the sensor spin
is in general non-stationary or device dependent~\cite{PaladinoRMP2014},
which means its correlation functions are not fully characterized
prior to the sensing experiment or vary in different runs of experiments.
Otherwise, the classical noise correlation can always be subtracted
from the correlation function, which is actually not feasible in realistic
experiments. For a rough estimation, we assume the uncertainty of the
classical noise correlation is in the same order of the correlation.

We consider a target spin at temperature $\gg\hbar\omega_{0}/k_{{\rm B}}$ and therefore has the
density operator $\hat{\rho}_{{\rm Q}}=1/2$.  Under this high-temperature condition, the second-order
correlation of the target
$
C_{{\rm Q}}^{++}(t_m,t_n)=a^{2}\cos\left[\omega_{0}(t_m-t_n)\right]
$.
The second-order correlation comes from both the quantum target and the classical noise
background 
\begin{align}
G_{yy}^{xx}(t_m,t_n)=\tau_{\rm I}^{2}\left[C_{{\rm Q}}^{++}(t_m,t_n)+C_{{\rm C}}^{++}(t_m,t_n)\right]+O\left(\tau_{\rm I}^{4}\right).
\label{eq_Gyy}
\end{align}
 The correlation of the target spin would be concealed if the uncertainty
of the classical correlation is greater than the target correlation
{[}Fig.~\ref{fig_c2c4}(a){]}. On the contrary, as discussed above,
the fourth-order correlation $G_{yzzy}^{xxxx}$ of the measurement
outputs can exclude the effects of the classical noise. The corresponding
fourth-order quantum correlation of the target is
\begin{align}
G_{yzzy}^{xxxx}(t_j,t_{k},t_m,t_n)=\tau_{\rm I}^{4} C_{\rm Q}^{+--+}(t_j,t_{k},t_m,t_n)+O\left(\tau_{\rm I}^{6}\right),
\label{eq_Gyzzy}
\end{align}
with $C_{\rm Q}^{+--+}(t_j,t_{k},t_m,t_n)=a^{4}\sin\left[\omega_{0}(t_j-t_{k})\right]\sin\left[\omega_{0}(t_{m}-t_{n})\right]$ for the spin-1/2 target.

\subsection{Effect of target decoherence}

Above we have assumed the target spin precesses ideally and therefore
its correlations oscillate without decay. Actually, if the correlation
$C^{++}_{\rm Q}(t_m,t_n)$ oscillates without decay, its Fourier transform,
i.e., the correlation spectrum, would present a $\delta$-peak at
frequency $\omega_{0}$, that is, the target resonance can be made
arbitrarily high and eventually above any uncertainty of background
noise spectrum by increasing the data acquisition time. Under realistic
conditions, however, the precession is always subjected to disturbance
and has a finite decay time. In turn, the target resonance is broadened.
A broadened target resonance has a finite height and therefore cannot
be resolved if the uncertainty of the background noise spectrum is
larger than the resonance height, no matter how long the data acquisition
time is.

There are two mechanisms of the target decoherence.

One is the intrinsic decoherence due to coupling between the target
and its environment. Usually, the transverse relaxation (decay of
the spin polarization in the $x$-$y$ plane, or the pure dephasing)
is much faster than the longitudinal relaxation. For the sake of simplicity,
we assume the intrinsic decoherence is characterized by a pure dephasing
rate $\gamma_{0}$.

Another mechanism is the quantum backaction due to the weak measurement
by the sensor. Between two recorded outputs at, e.g., $t_{n}$ and
$t_{m}$, there are ``idle'' measurements whose outputs are ``discarded''.
During an idle shot of measurement, the sensor can be regarded as a
``bath'' spin for the target. When the interaction time $\tau_{{\rm I}}$
is much shorter than the target precession period $2\pi/\omega_{0}$,
the effect of the target-sensor interaction and the
resultant entanglement during $\tau_{{\rm I}}$ amount to an instantaneous pure dephasing quantized along the $x$ axis for the target spin,
with a dephasing rate
\begin{equation}
\gamma_{{\rm M}}=\frac{1}{4\tau_{{\rm I}}}\sin^{2}\left(a\tau_{{\rm I}}\right).\label{eq:gamma_M}
\end{equation}
The strength of the weak measurement can be quantified by $\gamma_{{\rm M}}\tau_{{\rm I}}$
or simply, $\gamma_{{\rm M}}$.

Considering the intrinsic dephasing along the $z$ axis and the measurement-induced
dephasing along the $x$ axis, the target correlation functions become
\begin{align}
C_{{\rm Q}}^{++}(t_m,t_n) & =a^{2}\cos\left[\omega_{0}(t_m-t_n)\right]e^{-(\gamma_0+\gamma_{\rm M})(t_m-t_n)}, \nonumber \\
C_{{\rm Q}}^{+\text{\textendash}-+}(t_j,t_{k},t_m,t_n) & =a^{4}\sin\left[\omega_{0}(t_j-t_{k})\right]\sin\left[\omega_{0}(t_m-t_n)\right] \nonumber \\
 & \times e^{-(\gamma_0+\gamma_{\rm M})(t_j-t_{k})-2\gamma_{\rm M}(t_{k}-t_m)-(\gamma_0+\gamma_{\rm M})(t_m-t_n)}. \nonumber
\end{align}

\subsection{Effects of finite sensor-target interaction time}

Above we have assumed the sensor-target interaction time
for a shot of weak measurement $\tau_{\rm I}$ approaches to zero. Under realistic conditions,
$\tau_{\rm I}$ is always finite. The finiteness of the interaction time has two main effects on the detection sensitivity.

First, during a finite evolution time, the classical noise $B_{{\rm C}}$
will reduce the coherence of the sensor spin by a factor $L_{\rm C}$. If the interaction time
is not too long, i.e., $\left|B_{{\rm C}}\tau_{{\rm I}}\right|\lesssim1$ (which is usually the case),
the decoherence can be approximated as $L_{{\rm C}}\approx e^{-\frac{1}{2}\left\langle \phi_{{\rm C}}^{2}(t)\right\rangle }$
with $\phi_{{\rm C}}\equiv\int_{t}^{t+\tau_{{\rm I}}}B_{{\rm C}}(\tau)d\tau$.
For the measurement $xy$, the random noise
along the $z$ aixs will cause the measurement axis randomly deviate
from the $y$ direction and therefore reduce the measurement contrast
by a factor $L_{{\rm C}}$. On the contratry, for the measurement
$xz$, the measurement axis $z$ is not affected by the noise and
hence no reduction of contrast. As a result, the second-order correlation
$G_{yy}^{xx}$ and the fourth-order $G_{yzzy}^{xxxx}$,
both containing two measurements along the $y$ axis, will be reduced
by a factor of $L_{{\rm C}}^{2}$. See Appendix~\ref{Appendix_G2}
and~\ref{Appendix_G4} for the derivations.

Second, the sensor-target interaction during
a finite time results in quantum oscillation rather than an unbounded,
linearly increasing entanglement, therefore $\tau_{\rm I}$ in the prefactors in Eqs.~(\ref{eq_Gyy}) and (\ref{eq_Gyzzy}) is replaced with
$a^{-1}\sin(a\tau_{\rm I})\equiv a^{-1}\sin\alpha$ (see Appendix~\ref{Appendix_G2} for details).

Taking into account the effects of finite $\tau_{\rm I}$, the output
correlations of interest become
\begin{subequations}
\begin{align}
G_{yy}^{xx}\approx &  L_{{\rm C}}^{2} \sin^{2}\alpha\cos\left[\omega_{0}\left(t_m-t_n\right)\right]e^{-(\gamma_0+\gamma_{\rm M})(t_m-t_n)}
\nonumber \\
& +L_{{\rm C}}^{2}\left\langle \phi_{{\rm C}}\left(t_m\right)\phi_{{\rm C}}\left(t_n\right)\right\rangle ,
\label{subeq:G2}\\
G_{yzzy}^{xxxx}\approx & L_{{\rm C}}^{2}\sin^{4}\alpha\sin\left[\omega_{0}(t_j-t_{k})\right]\sin\left[\omega_{0}(t_m-t_n)\right]
\nonumber \\
&  \times e^{-(\gamma_0+\gamma_{\rm M})(t_j-t_{k})-2\gamma_{\rm M}(t_{k}-t_m)-(\gamma_0+\gamma_{\rm M})(t_m-t_n)}.
\label{subeq:G4}
\end{align}
\label{eq:correlation_signal}
\end{subequations}

\section{Sensing by different correlations - a comparison}

\subsection{Sensing by second-order correlation}

For the second-order correlation $G_{yy}^{xx}$, we use the sequence
of weak measurement shown in Fig. \ref{fig_measurement}(b). The output
of the $m$-th shot is $s_{m}^{xy}$, and the output correlation is
\begin{align*}
G_{yy}^{xx}\left(t_{m},t_{n}\right) & =\frac{1}{M}\sum_{j=0}^{M-m}s_{j+m}^{xy}s_{j+n}^{xy},
\end{align*}
 for $M\gg m$. Here $t_{j}\equiv j\tau_{{\rm I}}$. The correlation
spectrum is obtained by Fourier transform $\tilde{G}_{yy}^{xx}(\omega)=\sum_{n=0}^{N_{{\rm F}-1}}G_{yy}^{xx}\left(n\tau_{{\rm I}},0\right){e^{i\omega n\tau_{\mathrm{I}}}}$,
where $N_{{\rm F}}\tau_{{\rm I}}$ is the range of time for Fourier
transform. Using Eq. (\ref{subeq:G2}), we obtain the spectrum as
\[
\tilde{G}_{yy}^{xx}(\omega)=L_{\mathrm{C}}^{2}\left\{ \left[2\gamma_{{\rm M}}\tau_{{\rm I}}\frac{1-e^{-iN_{{\rm F}}\theta}}{1-e^{-i\theta}}+(\omega\rightarrow -\omega)^{*}\right]+\tau_{{\rm I}}{S_{\mathrm{C}}(\omega)}\right\} ,
\]
where $S_{\mathrm{C}}(\omega)\equiv \int_{-\infty}^{\infty}dt\langle B_{\mathrm{C}}(t)B_{\mathrm{C}}(0)\rangle e^{i\omega t}$ and $\theta\equiv (\omega-\omega_{0})\tau_{{\rm I}}-i(\gamma_{0}+\gamma_{{\rm M}})\tau_{{\rm I}}$.
Particularly at the target frequency $\omega=\omega_{0}$ , the spectrum
is
\[
\tilde{G}_{yy}^{xx}(\omega_{0})\approx L_{\mathrm{C}}^{2}\left[2\gamma_{{\rm M}}\tau_{{\rm I}}\frac{1-e^{-N_{{\rm F}}\tau_{{\rm I}}{(\gamma_{0}+\gamma_{{\rm M}})}}}{1-e^{-\tau_{{\rm I}}{(\gamma_{0}+\gamma_{{\rm M}})}}}+\tau_{{\rm I}}S_{\mathrm{C}}(\omega_{0})\right].
\]

\begin{figure}
\includegraphics[width=0.9\columnwidth]{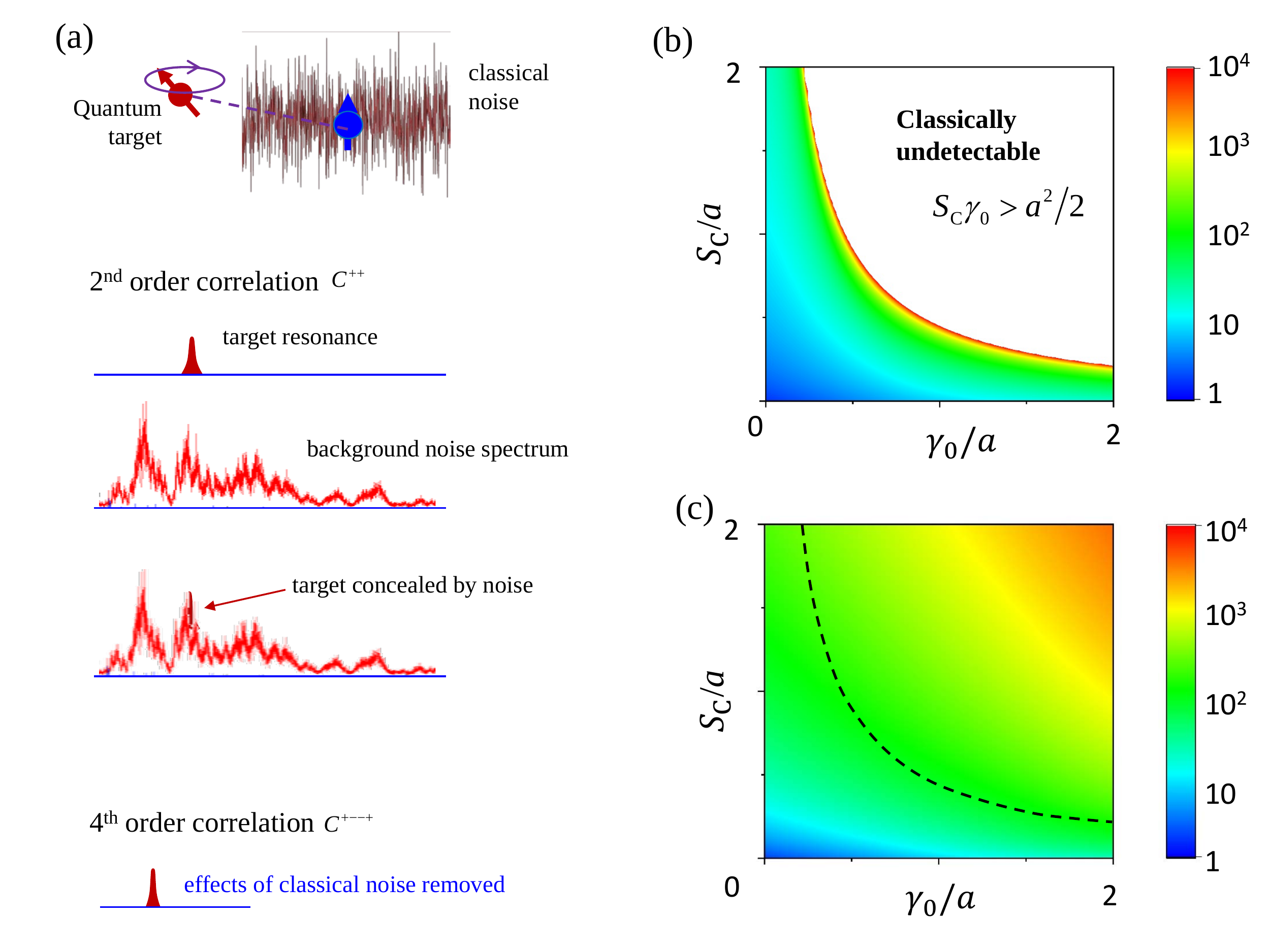}
\caption{(a). Illustration of classical-noise-free sensing via measuring quantum
correlation. Top: The sensor spin (blue arrow) is coupled to the target
spin-1/2 (purple arrow) and subjected to a non-stationary classical
noise. Middle: Spectra of the second-order classical correlations
of the target and the classical noise and their sum. The target signal
is concealed by the classical noise. Bottom: Spectrum of the high
order quantum correlation, to which the classical noise does not contribute.
(b) Optimal data acquisition time $T_{\mathrm{opt}}$ (in units of
$1/a$) for sensing a quantum target using the second-order classical
correlation, as a function of the classical noise strength $S_{{\rm C}}$
and the intrinsic dephasing rate $\gamma_{0}$ of the target spin
(in units of $a$). The white zone ($S_{{\rm C}}\gamma_{0}>a^{2}/2$)
is the parameter region where the quantum target is not detectable.
(c) Optimal data acquisition time for sensing a quantum target using
the fourth-order quantum correlation, as a function of the classical
noise strength and the intrinsic target dephasing rate resonance.
The dashed curve marks the condition $S_{\rm C}\gamma_0=a^2/2$.
\label{fig_c2c4}}
\end{figure}

The uncertainty of the output correlation has two sources -- the
shot noise $\sigma_{{\rm M}}$ of the total $M$ shots of measurement
and the uncertainty of the classical noise spectrum $\delta S_{\mathrm{C}}$
(which is assumed to be $\delta S_{{\rm C}}\sim S_{{\rm C}}$). The
total uncertainty of the correlation spectrum is
\[
\sigma=\sqrt{\sigma_{{\rm M}}^{2}+L_{\mathrm{C}}^{4}\tau_{{\rm I}}^{2}S_{\mathrm{C}}^{2}}.
\]
The shot noise of the measurement at the resonance frequency is $\sigma_{{\rm M}}=\sqrt{N_{{\rm F}}/M}$
if we assume that the readout fidelity of the sensor spin state is perfect.
While the shot noise increases with the range of transform ($N_{{\rm F}}$),
the target signal at the resonance frequency $\omega_{0}$ saturates
with $N_{{\rm F}}\tau_{{\rm I}}$ increasing beyond $T_{2}$ (since
the target spin would have no correlation beyond its coherence time).
To optimize the signal-to-noise ratio (SNR), we choose $N_{{\rm F}}\tau_{{\rm I}}\sim1/(\gamma_{0}+\gamma_{{\rm M}})$.
Under this condition,the strength of the signal is about $2\gamma_{{\rm M}}L_{\mathrm{C}}^{2}/(\gamma_{0}+\gamma_{{\rm M}})$
and hence the SNR becomes
\begin{align}
\mathrm{SNR}_{\mathrm{G}_{2}} & \equiv \frac{\tilde{G}_{yy}^{xx}(\omega_{0})}{\sigma}
\approx\frac{2\gamma_{{\rm M}}/(\gamma_{0}+\gamma_{{\rm M}})}{\sqrt{\sigma_{{\rm M}}^{2}+L_{\mathrm{C}}^{4}\tau_{{\rm I}}^{2}S_{\mathrm{C}}^{2}}}L_{\mathrm{C}}^{2}.
\nonumber
\end{align}
For a rough estimation, we assume that the classical noise has comparable
spectral density in the frequency range of interest. With this assumption,
the sensor decoherence due to the classical noise during the interaction
time $\tau_{{\rm I}}$ is $L_{{\rm C}}\approx e^{-\tau_{{\rm I}}S_{{\rm C}}/2}$.
With $\gamma_{\mathrm{M}}\approx\sin^{2}(a\tau_{{\rm I}})/(4\tau_{{\rm I}})\approx a^{2}\tau_{{\rm I}}/4$,
and hence $L_{\mathrm{C}}^{2}\sim e^{-4\gamma_{\mathrm{M}}S_{{\rm C}}/a^{2}}$,
we obtained the SNR as
\begin{equation}
\mathrm{SNR}_{\mathrm{G}_{2}}\approx\left[\frac{\gamma_{0}+\gamma_{{\rm M}}}{4\gamma_{{\rm M}}^{2}T}{e^{8\gamma_{\mathrm{M}}S_{{\rm C}}/a^{2}}}+\frac{4\left(\gamma_{0}+\gamma_{{\rm M}}\right)^{2}S_{{\rm C}}^{2}}{a^{4}}\right]^{-1/2},
\label{eq:G2_SNR}
\end{equation}
for a total data acquisition time $T=M\tau_{{\rm I}}$

The key issue of the second-order correlation sensing, as shown in Eq. (\ref{eq:G2_SNR}), is that the SNR is upper bounded by
\begin{equation}
{\rm SNR}_{\mathrm{G}_{2}}\le a^{2}/\left(2\gamma_{0}S_{{\rm C}}\right),
\end{equation}
no matter how long the data acquisition time $T$ is and how strong the measurement
back-action $\gamma_{\rm M}$ is. That is, when the combined classical noise spectral
density and target resonance width are greater than a threshold, namely,
$\gamma_{0}S_{{\rm C}}>a^{2}/2$ {[}the white zone in Fig.~\ref{fig_c2c4}(b){]},
the target is not detectable by the second-order correlation measurement.
Though in principle one can increase the
sensor-target coupling strength $a$ to increase the upper bound of the
SNR, there are always physical constraints on the coupling strength.
For example, the key parameters are related to
the magnetic moment of the sensor spin $\mu_{S}$ and that of the target spin $\mu_I$ via
$a\propto \mu_{S}\mu_I$, $\gamma_0\propto \mu_I^2$, and $S_{\rm C}\propto \mu_S^2$, so the threshold
$2\gamma_0S_{\rm C}/a^2$ is independent of the magnetic moments of the sensor and the target,
but is constrained by the environmental noise strengths and the target-sensor spatial configuration. That is, to overcome the
upper bound of SNR in sensing by second-order classical correlations, one has to either suppress the environmental noises
or place the sensor closer to the target or both.

To achieve a certain SNR, the data acquisition time $T$ can be worked
out from Eq. (\ref{eq:G2_SNR}) as a function of the measurement-induced
target spin relaxation $\gamma_{{\rm M}}$ (i.e., the measurement
strength). For ${\rm SNR}_{\mathrm{G}_{2}}=1$,
\[
T\approx\frac{\gamma_{0}+\gamma_{{\rm M}}}{4\gamma_{{\rm M}}^{2}}\frac{{e^{8\gamma_{\mathrm{M}}S_{{\rm C}}/a^{2}}}}{1-{4\left(\gamma_{0}+\gamma_{{\rm M}}\right)^{2}S_{{\rm C}}^{2}}/{a^{4}}}.
\]
As shown in Eq. (\ref{eq:gamma_M}),
the measurement strength $\gamma_{M}\equiv\sin^{2}(a\tau_{{\rm I}})/(4\tau_{{\rm I}})$
can be tuned by varying the interaction time $\tau_{{\rm I}}$ with
an upper bound $\gamma_{{\rm M}}\le\gamma_{{\rm M}}^{{\rm max}}\equiv a\max_{x}\frac{\sin^{2}x}{4x}\approx0.18a$.
We optimize the data acquisition time by tuning
$\gamma_{{\rm M}}$ for the combined noise strength and resonance
width below the threshold (i.e., $\gamma_{0}S_{{\rm C}}\le a^{2}/2$).
The result is plot in Fig.~\ref{fig_c2c4}(b).
{
The scaling relation between the optimal data acqusition time and the noise strength ($S_{\rm C}$), the target dephasing rate ($\gamma_0$), and the sensor-target coupling ($a$ or $\gamma_{\rm M}^{\rm max}\approx 0.18 a$)
can be approximated as
\begin{align}
T_{\rm opt}^{\rm 2nd} \sim
\begin{cases}
\frac{S_{\rm C}}{a^2} \left(1-\frac{2\gamma_0 S_{\rm C}}{a^2}\right)^{-3},  &  \left(\gamma_0+\gamma^{\rm max}_{\rm M}\right)S_{\rm C} \gtrsim {a^2}/{2}, \\
\frac{1}{\gamma_{\rm M}^{\rm max}} \frac{1+\gamma_0/\gamma_{\rm M}^{\rm max}}{1-2\gamma_0S_{\rm C}/{a^2}}, & {\left(\gamma_0+\gamma^{\rm max}_{\rm M}\right)S_{\rm C}} \lesssim  {a^2}/{2} ,
\end{cases}
\nonumber
\end{align}
up to a factor $\sim O\left( 1\right)$, for the strong and weak noise conditions, respectively.
See Appendix~\ref{Append_sec_optimal_time_2nd} for the derivation. Note that the time diverges when the parameters approach to the threshold $2\gamma_0S_{\rm C}/a^2=1$.
}

\subsection{Classical-noise-free sensing by fourth-order quantum correlation}

For the fourth-order correlation $G_{yzzy}^{xxxx}$, we use the sequence
of weak measurement shown in Fig.~\ref{fig_measurement}(c), where
the $m$-th cycle contains two shots of measurement labeled as $xy$
and $xz$, with outputs $s_{m}^{xy}$ and $s_{m}^{xz}$, respectively.
The output correlation is
\[
G_{yzzy}^{xxxx}\left(t_{j},t_{k},t_{m},t_{n}\right)\approx\frac{1}{M}\sum_{i=0}^{M-{j}}s_{i+j}^{xy}s_{i+k}^{xz}s_{i+m}^{xz}s_{i+n}^{xy},
\]
 for $M\gg{j}$. Here $t_{j}\equiv2j\tau_{{\rm I}}$ .
By three-dimensional Fourier transform of the fourth-order correlation $G_{yzzy}^{xxxx}$
in Eq. (\ref{subeq:G4}), the signal of the target
spin at the resonance frequency is obtained as
\[
\tilde{G}_{yzzy}^{xxxx}(\omega_{0},0,\omega_{0})\approx L_{\mathrm{C}}^{2}{\left(\frac{\gamma_{\mathrm{M}}}{\gamma_{\mathrm{M}}+\gamma_{0}}\right)^{2}\frac{1}{4\gamma_{\mathrm{M}}\tau_{{\rm I}}}}.
\]

In contrast to the second-order signal $\tilde{G}_{yy}^{xx}$, the classical
noise background is absent. The shot noise in the frequency domain
is 
\[
\sigma_{{\rm M}}=\frac{\sqrt{N_{\mathrm{F},2}}\sqrt{N_{\mathrm{F},1}}\sqrt{N_{\mathrm{F},2}}}{\sqrt{M}}=N_{\mathrm{F},2}\sqrt{N_{\mathrm{F},1}}\sqrt{2\tau_{{\rm I}}/T}.
\]
Here,  for optimal SNR,  the number of data points taken
 in Fourier transform
for $t_{j}-t_{k}$ and $t_{m}-t_{n}$  is $N_{\mathrm{F},2}\approx1/\left[{2}(\gamma_{0}+\gamma_{{\rm M}})\tau_{{\rm I}}\right]$ , and $N_{\mathrm{F},1}\approx1/(4\gamma_{{\rm M}}\tau_{{\rm I}})$
for $t_{k}-t_{m}$. The total data acquisition time is $T\approx2M\tau_{{\rm I}}$.
The SNR is
\[
\mathrm{SNR}_{\mathrm{G}_{4}}\equiv \frac{\tilde{G}_{yzzy}^{xxxx}(\omega_{0},0,\omega_{0})}{\sigma_{\rm M}}
\approx{\frac{1}{\sqrt{2}}}\frac{\gamma_{\mathrm{M}}^{3/2}\sqrt{T}}{\gamma_{\mathrm{M}}+\gamma_{0}}e^{-4\gamma_{\mathrm{M}}S_{{\rm C}}/a^{2}},
\]
 The data acquisition time required to detect the target ($\mathrm{SNR_{\mathrm{G}_{4}}}>1$) is
\[
{T=2\frac{\left(\gamma_{\mathrm{M}}+\gamma_{0}\right)^{2}}{\gamma_{\mathrm{M}}^{3}}e^{8\gamma_{\mathrm{M}}S_{{\rm C}}/a^{2}}}.
\]
It can be optimized by tuning the measurement strength in the range $0\le\gamma_{\mathrm{M}}\le\gamma_{\mathrm{M}}^{\mathrm{max}}$. The result is plot in Fig.~\ref{fig_c2c4}(c).
{ The approximate
scaling relations between the optimal data acquisition time and the classical noise strength, the target dephasing rate, and the sensor-target coupling strength is
\[
T_{\mathrm{opt}}^{\rm 4th} \sim
\begin{cases}
\frac{8S_{\rm C}}{a^2}\left(1+\frac{8\gamma_0S_{\mathrm{C}}}{a^2}\right)^{2} , & \gamma_{\rm M}^{\rm max}S_{{\rm C}} \gtrsim a^2/8,\\
\frac{1}{\gamma_{\rm M}^{\rm max}} \left(1+\frac{\gamma_0}{\gamma_{\rm M}^{\rm max}}\right)^{2},  & \gamma_{\rm M}^{\rm max}S_{{\rm C}} \lesssim a^2/8,
\end{cases}
\]
up to a factor $\sim O(1)$, for the strong and weak noise cases, respectively. See Appendix~\ref{Append_sec_optimal_time_4th} for the derivation.}
 In contrast to the second-order
correlation approach, the fourth-order quantum correlation can always
have enough SNR by increasing the data acquisition time no matter
how strong the classical noise is.

\section{Conclusion and discussion}

Using the example of sensing a single spin, we show that the quantum
correlations of a target can be employed to enable classical-noise-free sensing
schemes. When the noise has strong non-stationary fluctuations in its correlation spectrum,
it would be impossible to detect a target by conventional correlation spectroscopy
that measures correlations of classical nature. On the contrary, quantum correlations can be measured to fully
exclude the effects of the classical noise so that the quantum object
is detected. As compared with the conventional noise filtering schemes, the higher-order quantum
correlation sensing does not depend on the specific properties of the classical noises,
be it strong or weak, slow or fast, and Gaussian or non-Gaussian.

We would like to remark on when a noise can be regarded as $\emph{classical}$,
since, after all, all objects interacting with a sensor are ultimately
quantum. In fact, if there are many ($N\gg1$) particles interacting
weakly with a sensor, with coupling to each individual particle scaling
as $a\sim 1/\sqrt{N}$, the interaction between the sensor would induce
negligible back-action on the particles at the macroscopic limit $N\rightarrow\infty$.
At this limit, the fourth-order quantum correlation ($\sim N\times a^4\sim N^{-1}$) would become vanishingly small
relative to the second-order classical correlation ($\sim N\times a^2\sim O(1)$). Thus, consistent with our intuition, the
coupling to a macroscopic object (e.g., a magnet that supplies a \textquotedblleft classical\textquotedblright{}
field) can be regarded as classical and its instability (due to, e.g., temperature fluctuation) regarded as a classical noise.

Measurement of the quantum correlations is of interest in studying
quantum many-body physics at mesoscopic scales. The conventional measurement
involves classical probes and therefore cannot detect the quantum
correlations. Quantum sensing of quantum correlations may reveal new
characteristics of quantum many-body systems (such as quantum entanglement,
correlations violating Leggett-Garg inequalities, and topological
orders).

\begin{acknowledgments}
This work was supported by Hong Kong RGC/GRF Project 14300119.
\end{acknowledgments}

\begin{appendix}

\section{Derivation of $G^{xx}_{yy}$}
\label{Appendix_G2}

In each shot of measurement $xy$, the initial state of the sensor is
$$|x\rangle\equiv \frac{|+\rangle+|-\rangle}{\sqrt{2}},$$
where $|\pm\rangle$ are the sensor eigenstates of $\hat{\sigma}_z$ with eigenvalues $\pm 1$.
The Hamiltonian in the Schr\"{o}dinger picture can be rewritten as
\[
\hat{H}(t)=\sum_{{s=\pm1}}\left[\omega_{0}\hat{I}_{z}+as\hat{I}_{x}+\frac{1}{2}sB_{{\rm C}}(t)\right]\otimes|{s}\rangle\langle{s}|,
\]
The evolution from $t_m$ to $t_m+\tau_{\rm I}$ can be obtained as
$$
\hat{U}_m=
e^{-i\phi_m/2}\hat{U}_+\otimes |+\rangle \langle +|+
e^{+i\phi_m/2}\hat{U}_-\otimes |-\rangle\langle -|,
$$
where $\phi_m\equiv \int_{t_m}^{t_m+\tau_{\rm I}} B_{\rm C}(t) dt$ is the phase shift due to the classical noise, and
$$
\hat{U}_{\pm}\equiv  e^{-i\left(\omega_0\hat {I}_z\pm a \hat{I}_x\right)\tau_{\rm I}},
$$
is the target state evolution conditioned on the sensor state. If the interaction time is short, i.e., $a\tau_{\rm I}\ll1 $  and $\omega_0\tau_{\rm I} \ll 1$, the conditional evolution can be
approximated as~\cite{LiuPRL2017}
\begin{align}
\hat{U}_{\pm}\approx e^{-i\varphi\hat{I}_{\mathrm{z}}}e^{\mp i\alpha\hat{I}_{x}},
\label{eq:short_time}
\end{align}
with $\varphi=\omega_{0}\tau_{\rm I}$ and $\alpha=a\tau_{\rm I}$~\cite{MaLiuPRAPPLIED2016,PfenderNC2019}. For longer evolution time, such decomposition is also possible, but $\alpha$ and the $x$
direction would be modified~\cite{MaLiuPRAPPLIED2016,PfenderNC2019}.

For convenience, we introduce four basic super-operators constructed
by $\hat{U}_{\pm}$
\begin{subequations}
\begin{align}
\hat{\mathcal M}_{x}\hat{\rho}_{\mathrm{Q}} & =\mathrm{Tr}_{\mathrm{S}}\left[\hat{\sigma}_{x}\hat{U}\hat{\rho}_{\mathrm{Q}}\otimes|x\rangle\langle x|\hat{U}^{\dagger}\right]=\frac{\hat{U}_{-}\hat{\rho}_{\mathrm{Q}}\hat{U}_{+}^{\dagger}+{\rm h.c}}{2}, \\
\hat{\mathcal M}_{y}\hat{\rho}_{\mathrm{Q}} & =\mathrm{Tr}_{\mathrm{S}}\left[\hat{\sigma}_{y}\hat{U}\hat{\rho}_{\mathrm{Q}}\otimes|x\rangle\langle x|\hat{U}^{\dagger}\right]=\frac{\hat{U}_{-}\hat{\rho}_{\mathrm{Q}}\hat{U}_{+}^{\dagger}-{\rm h.c}}{2i}, \\
\hat{\mathcal M}_{z}\hat{\rho}_{\mathrm{Q}} & =\mathrm{Tr}_{\mathrm{S}}\left[\hat{\sigma}_{z}\hat{U}\hat{\rho}_{\mathrm{Q}}\otimes|x\rangle\langle x|\hat{U}^{\dagger}\right]=\frac{\hat{U}_{+}\hat{\rho}_{\mathrm{Q}}\hat{U}_{+}^{\dagger}-\hat{U}_{-}\hat{\rho}_{\mathrm{Q}}\hat{U}_{-}^{\dagger}}{2},\\
\hat{\mathcal M}_{0}\hat{\rho}_{\mathrm{Q}} &=\mathrm{Tr}_{\mathrm{S}}\left[\hat{U}\hat{\rho}_{\mathrm{Q}}\otimes|x\rangle\langle x|\hat{U}^{\dagger}\right]=\frac{\hat{U}_{+}\hat{\rho}_{\mathrm{Q}}\hat{U}_{+}^{\dagger}+\hat{U}_{-}\hat{\rho}_{\mathrm{Q}}\hat{U}_{-}^{\dagger}}{2},
\end{align}
\label{eq:super}
\end{subequations}
with $\hat{U}\equiv \hat{U}_+\otimes |+\rangle\langle +|+\hat{U}_-\otimes |-\rangle\langle -|$.
Using Eq.~(\ref{eq:short_time}), we work out the four basic super-operators explicitly as
\begin{subequations}
\begin{align}
\hat{\mathcal M}_{0} & =\hat{\mathcal U}_{\varphi}\left[\cos\alpha+2\sin^{2}\frac{\alpha}{2}\left(2\hat{\mathcal I}_{x}^{+}\right)^{2}\right], \\
\hat{\mathcal M}_{x} & =\hat{\mathcal U}_{\varphi}\left[1-2\sin^{2}\frac{\alpha}{2}\left(2\hat{\mathcal I}_{x}^{+}\right)^{2}\right], \\
\hat{\mathcal M}_{y} & =\hat{\mathcal U}_{\varphi}\left(2\hat{\mathcal I}_{x}^{+}\right)\sin\alpha, \\
\hat{\mathcal M}_{z} & =\hat{\mathcal U}_{\varphi}\left(2\hat{\mathcal I}_{x}^{-}\right)\sin\alpha,
\end{align}
\label{eq:super_single}
\end{subequations}
where $\hat{\mathcal U}_{\varphi}\hat{\rho}_{\rm Q}\equiv e^{-i\varphi\hat{I}_{\mathrm{z}}}\hat{\rho}_{\rm Q}e^{i\varphi\hat{I}_{\mathrm{z}}}$ is the free precession about the $z$-axis during $\tau_{\rm I}$.
The following rules are useful for calculation and for understanding the physical meanings
\begin{subequations}
\begin{align}
2\hat{\mathcal I}_{x}^{+}\left(\hat{1}\right) & =\mathbf{e}_{x}\cdot\hat{\boldsymbol{\sigma}}, \\
2\hat{\mathcal I}_{x}^{+}\left(\mathbf{n}\cdot\hat{\boldsymbol{\sigma}}\right) & =\mathbf{e}_{x}\cdot\mathbf{n}\hat{1}, \\
2\hat{\mathcal I}_{x}^{-}\left(\hat{1}\right) & =0, \\
2\hat{\mathcal I}_{x}^{-}\left(\mathbf{n}\cdot\hat{\boldsymbol{\sigma}}\right)& =\left(\mathbf{e}_{x}\times\mathbf{n}\right)\cdot\hat{\boldsymbol{\sigma}}.
\end{align}
\label{eq:rules}
\end{subequations}
{The physical meaning of the super-operators $\hat{\mathcal M}_{x/y/z}$ are: The sensor spin is initially prepared in the state $|x\rangle$ and
later measured along the $x/y/z$ basis, and the difference between the target states for the two outputs $\pm 1$ is given by the super-operator acting on the initial target state.
In particular, the measurement $xy$ (corresponding to $\hat{\mathcal M}_y$) would induce weak polarization of the target spin along the $x$ axis. The measurement $xz$ (corresponding to $\hat{\mathcal M}_z$) would induce
a rotation about the $x$ axis, which can lift the spin polarization away from the $xy$ plane.
The superoprator $\hat{\mathcal M}_0$ can be written as $\hat{\mathcal U}_{\varphi}\hat{\mathcal L}_x$ with $\hat{\mathcal L}_x\equiv \cos\alpha+2\sin^{2}\frac{\alpha}{2}\left(2\hat{\mathcal I}_{x}^{+}\right)^{2}$ being the Lindblad
super-operator for pure dephasing along the $x$ axis. It has a clear physical meaning: the target spin, during the free precession about the $z$ axis, is subjected to a pure dephasing along the $x$ axis due to the quantum back-action of the
weak measurrement. Actually, when the measurment result of the sensor is ``unknown'' or discarded, the sensor acts as an ``environment'' coupled to the $x$-component of the target spin, so the effect is dephasing of the target along the $x$ axis.}

The projective measurement of the sensor operator $\hat{\sigma}_y$ has two outputs $s^{xy}_m=\pm 1$ for the two eigenstate $|\pm y\rangle$,
with the corresponding target states (not normalized) after the measurement being
\begin{align}
\hat{\rho}^{\pm}_{\rm Q}(t_{m+1}) & =\langle \pm y|\hat{U}_m|x\rangle \hat{\rho}_{\rm Q}(t_m) \langle x|\hat{U}_m^{\dag}|\pm y\rangle
\nonumber \\
& \equiv \hat{M}_m^{\pm}\hat{\rho}_{\rm Q}(t_m) \left(\hat{M}_m^{\pm}\right)^{\dag}
\nonumber \\
 & \equiv \hat{\mathcal M}_m^{\pm}\hat{\rho}_{\rm Q}(t_m), \nonumber
\end{align}
where the Kraus operators are defined as  $\hat{M}_m^{\pm} \equiv \langle \pm y|\hat{U}_m|x\rangle$, and
the superoperator
$ \hat{\mathcal M}_m^{\pm}\hat{A}\equiv  \hat{M}_m^{\pm}\hat{A} \left(\hat{M}_m^{\pm}\right)^{\dag}$.
The probability of the output $s^{xy}_m=\pm 1$ is
$$ p^{\pm}_m =\mathrm{Tr}\hat{\rho}^{\pm}_{\rm Q}(t_{m+1})=
{\rm Tr} \hat{\mathcal M}_m^{\pm}\hat{\rho}_{\rm Q}(t_m). $$
If the interaction is weak and/or the interaction time is short ($a\tau_{\rm I}\ll 1$),
the target states after the measurement for the two different outputs would be only weakly distinguishable, hence weak measurement.

If the measurement output is discarded, the target state after the measurement would be
\[
\hat{\rho}_{{\rm Q}}(t_{m+1}{)}=\hat{\mathcal{M}}_{m}^{+}\hat{\rho}_{{\rm Q}}(t_{m})+\hat{\mathcal{M}}_{m}^{-}\hat{\rho}_{{\rm Q}}(t_{m})\equiv\hat{\mathcal{M}}_{m}\hat{\rho}_{{\rm Q}}(t_{m}),
\]
where $\hat{\mathcal M}_m\equiv \hat{\mathcal M}_m^{+}+\hat{\mathcal M}_m^{-}$. The superoperator $\hat{\mathcal M}_m$ can be worked out explicitly as
\begin{align}
\hat{\mathcal M}_m\hat{\rho}_{\rm Q} & ={\rm Tr}_{\rm S} \hat{U}_m\hat{\rho}_{\rm Q}\otimes |x\rangle\langle x| \hat{U}_m^{\dag} \nonumber \\
 &= \frac{\hat{U}_+\hat{\rho}_{\rm Q}\hat{U}_+^{\dag}+\hat{U}_-\hat{\rho}_{\rm Q}\hat{U}_-^{\dag}}{2}=\hat{\mathcal M}_0\hat{\rho}_{\rm Q},
\nonumber
\end{align}
where ${\rm Tr}_{\rm S}$ denoting the partial trace of the sensor.
As discussed above, the super-operator $\hat{\mathcal M}_0$ corresponds to a measurement-induced dephasing along the $x$ axis.
If the interaction time is short as compared with the target spin precession period, i.e., $\omega_0\tau_{\rm I}\ll 1$, the dephasing can be regarded as instantaneous with a rate
$$
\gamma_{\rm M}=\frac{1}{4\tau_{\rm I}}\sin^2(a\tau_{\rm I}).
$$
The strength of the weak measurement is quantified by $\gamma_{\rm M}\tau_{\rm I}$.

The joint probability of two outcomes of measurements separated by $n$ shots is
$$p\left(u_{m},u_{m+n}\right)=\mathrm{Tr}\left[\hat{\mathcal M}^{u_{m+n}}_{m+n}\hat{\mathcal M}_0^{n-1}\hat{\mathcal M}^{u_{m}}_{m}\hat{\rho}_{B}\right].$$
The second-order correlation between the outputs is
$$G_{yy}^{xx}(t_{m+n},t_{m})=\left\langle \sum_{u_{m}u_{m+n}}u_{m}u_{m+n}p(u_{m},u_{m+n})\right\rangle .$$
 $\langle \cdots\rangle$
denotes the ensemble average of the classical noise $B_{\mathrm{C}}(t)$.
Using the super-operators, the correlation can also be reformulated
as
\begin{equation}
G_{yy}^{xx}(t_{m+n},t_{m})=\left\langle \mathrm{Tr}\left[\hat{\mathcal M}_{y,m+n}\hat{\mathcal M}_0^{n-1}\hat{\mathcal M}_{y,m}\hat{\rho}_{\mathrm{Q}}\right]\right\rangle \label{eq:2ndclassical}
\end{equation}
where $\hat{\mathcal M}_{y,m}=\hat{\mathcal M}^{+}_{m}-\hat{\mathcal M}^{-}_{m}$~\cite{PfenderNC2019}. The superoperator $\hat{\mathcal M}^{xy}_m$
can be expanded as $\hat{\mathcal M}_{y,m}=\sin\phi_{m}\hat{\mathcal M}_{x}+\cos\phi_{m}\hat{\mathcal M}_{y}$ {(with the physical meaning that the classical noise rotates the measurement axis of the target spin
about the $z$ axis by a random angle $\phi_m$)}.

To consider the intrinsic dephasing of the target
spin along the $z$ axis, we introduce the pure dephasing Lindblad
super-operator
\[
\hat{\mathcal{L}}_{z}\hat{\rho}\equiv\left[e^{-{2}\gamma_{0}\tau_{{\rm I}}}+\left(1-e^{-{2}\gamma_{0}\tau_{{\rm I}}}\right)\left(2\hat{\mathcal{I}}_{z}^{+}\right)^{2}\right]\hat{\rho}
\]
The evolution of the target spin under idle measurement
$\hat{\mathcal{M}}_{0}$ should be modified to be
\[
\hat{\mathcal{M}}=\hat{\mathcal{M}}_{0}\hat{\mathcal{L}}_{z}=\hat{\mathcal{U}}_{0}\hat{\mathcal{L}}_{x}\hat{\mathcal{L}}_{z}.
\]
Therefore, the second-order correlation in Eq.~(\ref{eq:2ndclassical}) can be expanded as
\begin{align}
G_{yy}^{xx}(t_{m+n},t_{m}) & =\langle\sin\phi_{m+n}\sin\phi_{m}\rangle\mathrm{Tr}\left[\hat{\mathcal M}_{x}\hat{\mathcal M}^{n-1}\hat{\mathcal M}_{x}\hat{\rho}_{\mathrm{Q}}\right] \nonumber \\
&+\langle\sin\phi_{m+n}\cos\phi_{m}\rangle\mathrm{Tr}\left[\hat{\mathcal M}_{x}\hat{\mathcal M}^{n-1}\hat{\mathcal M}_{y}\hat{\rho}_{\mathrm{Q}}\right] \nonumber \\
&+\langle\cos\phi_{m+n}\sin\phi_{m}\rangle\mathrm{Tr}\left[\hat{\mathcal M}_{y}\hat{\mathcal M}^{n-1}\hat{\mathcal M}_{x}\hat{\rho}_{\mathrm{Q}}\right] \nonumber\\
&+\langle\cos\phi_{m+n}\cos\phi_{m}\rangle\mathrm{Tr}\left[\hat{\mathcal M}_{y}\hat{\mathcal M}^{n-1}\hat{\mathcal M}_{y}\hat{\rho}_{\mathrm{Q}}\right].
\label{eq:G2weak}
\end{align}
Using the properties of these super-operators in Eq.~(\ref{eq:rules}),
we have $\hat{\mathcal M}_{x}\hat{1}=\cos\alpha\hat{1}$,
$\hat{\mathcal M}\hat{1}=\hat{1}$, and therefore
$
\mathrm{Tr}\left[\hat{\mathcal M}_{y}\hat{\mathcal M}^{n-1}\hat{\mathcal M}_{x}\hat{\rho}_{\mathrm{Q}}\right]=0
$ for $\hat{\rho}_{\rm Q}=\hat{1}/2$. Similarly,
$\mathrm{Tr}\left[\hat{\mathcal M}_{x}\hat{\mathcal M}^{n-1}\hat{\mathcal M}_{y}\hat{\rho}_{\mathrm{Q}}\right]=0$, $\mathrm{Tr}\left[\hat{\mathcal M}_{x}\hat{\mathcal M}^{n-1}\hat{\mathcal M}_{x}\hat{\rho}_{\mathrm{Q}}\right]=\cos^2\alpha$, and
\begin{align}
\mathrm{Tr}\left[\hat{\mathcal M}_{y}\hat{\mathcal M}^{n-1}\hat{\mathcal M}_{y}\hat{\rho}_{\mathrm{Q}}\right] &
= \sin\alpha\mathrm{Tr}\left[\hat{\mathcal M}_{y}\hat{\mathcal M}^{n-1}\hat{\mathcal U}_{\varphi}\left(\mathbf{e}_{x}\cdot\hat{\boldsymbol{\sigma}}/2\right)\right]
 \nonumber \\ &
\approx
\sin^{2}\alpha\cos (n\varphi){e^{-\frac{n-1}{4}\sin^{2}\alpha}}{\times e^{-(n-1)\gamma_0\tau_{\rm I}}}.
\label{eq:f2}
\end{align}

The correlations of the classical noise are
\begin{subequations}
\begin{align}
\langle\cos\phi_{m+n}\cos\phi_{m}\rangle
& \approx  e^{-\langle\phi_{m}^{2}\rangle/2-\langle\phi^2_{m+n}\rangle/2}\cosh\left\langle \phi_{m+n}\phi_{m}\right\rangle
\nonumber \\ & \approx  L_{\rm C}^2\cosh\left\langle \phi_{m+n}\phi_{m}\right\rangle, \\
\langle\sin\phi_{m+n}\sin\phi_{m}\rangle
& \approx  e^{-\langle\phi_{m}^{2}\rangle/2-\langle\phi^2_{m+n}\rangle/2}\sinh\left\langle \phi_{m+n}\phi_{m}\right\rangle
\nonumber \\ & \approx  L_{\rm C}^2\sinh\left\langle \phi_{m+n}\phi_{m}\right\rangle,
\end{align}
\label{eq:f4}
\end{subequations}
under the condition $|\phi_{m}|\ll 1$.

Inserting the results of Eqs.~(\ref{eq:f2}) and (\ref{eq:f4}) into Eq.~(\ref{eq:G2weak}), we obtain the second-order correlation $G_{yy}^{xx}(t_{m+n},t_{m})$ as in Eq.~(\ref{subeq:G2}) of the main text.

\section{Derivation of $G_{yzzy}^{xxxx}$}
\label{Appendix_G4}

{In the measurement of the fourth-order correlation $G^{xxxx}_{yzzy}$, each cycle consists of two shots of measurement $xy$ and $xz$, as shown in Fig.~\ref{fig_measurement}(c).
The super-operator for the measurement $xz$ as
\begin{align}
\hat{\mathcal M}_{z,m}\hat{\rho}_{\rm Q}  & \equiv \langle +|\hat{U}_m  |x\rangle \hat{\rho}_{\rm Q}\langle x|\hat{U}^{\dag}_{m}|+\rangle
-  \langle -|\hat{U}_m  |x\rangle \hat{\rho}_{\rm Q}\langle x|\hat{U}^{\dag}_{m}|-\rangle\nonumber \\
& =\frac{\hat{U}_+ \hat{\rho}_{\rm Q}\hat{U}^{\dag}_+-\hat{U}_- \hat{\rho}_{\rm Q}\hat{U}^{\dag}_{-}}{2} =\hat{\mathcal M}_z\hat{\rho}_{\rm Q}.
\end{align}
Unlike the measurement $xy$, the super-operator $\hat{\mathcal M}^{xz}_m$ is independent of $m$, i.e., not affected by the classical noise. Physically, this is because the noise along the $z$-axis has not effect on the
measurement axis $z$.}

The fourth-order correlation $G_{yzzy}^{xxxx}$, for four shots of measurement $xy$, $xz$, $xz$, and $xy$ at $t_j$, $t_k$, $t_m$, and $t_n$, correspondingly, is
\begin{subequations}
\begin{align}
& G_{yzzy}^{xxxx}(t_j,t_{k},t_m,t_n) \nonumber  \\
 = & \left\langle \mathrm{Tr}\left[\hat{\mathcal M}_{y,j}\hat{\mathcal M}^{2(j-k-1)}\hat{\mathcal M}_{z,k}\hat{\mathcal M}^{2(k-m-1)}\hat{\mathcal M}_{z,m}\hat{\mathcal M}^{2(m-n-1)}
\hat{\mathcal M}_{y,n}\hat{\rho}_{\mathrm{Q}}\right]\right\rangle
\nonumber \\
=&
 \left\langle \cos\phi_{j}\cos\phi_{n}\right\rangle
 \mathrm{Tr}\left[\hat{\mathcal M}_{y}\hat{\mathcal M}^{{2(j-k-1)}}\hat{\mathcal M}_{z}\hat{\mathcal M}^{{2(k-m-1)}}
\hat{\mathcal M}_{z}\hat{\mathcal M}^{{2(m-n-1)}}\hat{\mathcal M}_{y}\hat{\rho}_{\mathrm{Q}}\right]
\label{eq:G4a} \\ + &
\left\langle \sin\phi_{j}\sin\phi_{n}\right\rangle
\mathrm{Tr}\left[\hat{\mathcal M}_{x}\hat{\mathcal M}^{{2(j-k-1)}}\hat{\mathcal M}_{z}\hat{\mathcal M}^{{2(k-m-1)}}
\hat{\mathcal M}_{z}\hat{\mathcal M}^{{2(m-n-1)}}\hat{\mathcal M}_{x}
\hat{\rho}_{\mathrm{Q}}\right]
\label{eq:G4b} \\ + &
\left\langle \sin\phi_{j}\cos\phi_{n}\right\rangle
\mathrm{Tr}\left[\hat{\mathcal M}_{x}\hat{\mathcal M}^{{2(j-k-1)}}\hat{\mathcal M}_{z}
\hat{\mathcal M}^{{2(k-m-1)}}\hat{\mathcal M}_{z}\hat{\mathcal M}^{{2(m-n-1)}}
\hat{\mathcal M}_{y}\hat{\rho}_{\mathrm{Q}}\right]
\label{eq:G4c} \\  +&
\left\langle \cos\phi_{j}\sin\phi_{n}\right\rangle
\mathrm{Tr}\left[\hat{\mathcal M}_{y}\hat{\mathcal M}^{{2(j-k-1)}}\hat{\mathcal M}_{z}
\hat{\mathcal M}^{{2(k-m-1)}}\hat{\mathcal M}_{z}\hat{\mathcal M}^{{2(m-n-1)}}\hat{\mathcal M}_{x}\hat{\rho}_{\mathrm{Q}}\right].
\label{eq:G4d}
\end{align}
\end{subequations}
Using relations in Eq.~(\ref{eq:super_single}) and Eq.~(\ref{eq:rules}), it is easy to show that all the terms in Eqs.~(\ref{eq:G4b}), (\ref{eq:G4c}) and (\ref{eq:G4d}) vanish and only the term in (\ref{eq:G4a}) is non-zero. The evaluation of the first term~\cite{PfenderNC2019}
yields
\begin{align}
& G_{yzzy}^{xxxx}(t_{j},t_{k},t_{m},t_{n})
\nonumber \\   &  \approx   L_{{\rm C}}^{2}\sin^{4}\alpha\sin\left[{2}(j-k)\varphi\right]\sin\left[{2}(m-n)\varphi\right]
\nonumber \\ &
\times e^{-{2(j-k-1)\left(\frac{1}{4}\sin^{2}\alpha+\gamma_{0}\tau_{{\rm I}}\right)}}e^{-{2(k-m-1)}\times2\sin^{2}(\alpha/2)}e^{-{2(m-n-1)\left(\frac{1}{4}\sin^{2}\alpha+\gamma_{0}\tau_{{\rm I}}\right)}},
\nonumber
\end{align}
which is just Eq. (\ref{subeq:G4}) in the main text
with the substitution
$\gamma_{{\rm M}}=\frac{1}{4\tau_{{\rm I}}}\sin^{2}\alpha$
and $\varphi=\omega_{0}\tau_{{\rm I}}$ for $\alpha\lesssim1$.

{
\section{Derivation of optimal data acquisition times}
\subsection{Optimal time for sensing based on second-order correlations}
\label{Append_sec_optimal_time_2nd}

We define the dimensionless dephasing rates of the target spin as
\begin{align}
\bar{\gamma}_0 & \equiv \gamma_0 2S_{\rm C}/a^2, \nonumber \\
\bar{\gamma}_{\rm M} & \equiv \gamma_{\rm M}2S_{\rm C}/a^2, \nonumber
\end{align}
and write the data acquisition time  for the second-order correlation sensing as
$$
T=\frac{S_{\rm C}}{2a^2}\frac{\left(\bar{\gamma}_0+\bar{\gamma}_{\rm M}\right) e^{4\bar{\gamma}_{\rm M}}}{\bar{\gamma}_{\rm M}^2\left[1-\left(\bar{\gamma}_0+\bar{\gamma}_{\rm M}\right)^2\right]}.
$$
By this formula, the optimal time depends essentially only on one parameter, namely, $\bar{\gamma}_0$, except that the optimal $\bar{\gamma}_{\rm M}$ has to be checked against the physical constraint $\gamma_{\rm M}\le \gamma_{\rm M}^{\rm max}$.
Note that the parameters should be within the ranges $\bar{\gamma}_0< 1$ and $0<\bar{\gamma}_{\rm M} < 1-\bar{\gamma}_0$, so
$1+\left(\bar{\gamma}_0+\bar{\gamma}_{\rm M}\right)\sim O(1)$ and $e^{4\bar{\gamma}_{\rm M}} \sim O(1)$. Dropping the factors $\sim O(1)$ for the sake of simplicity in a rough estimation,
we get
$$
T^{-1}
\sim \frac{a^2}{S_{\rm C}}\frac{\bar{\gamma}_{\rm M}^2\left(1-\bar{\gamma}_0-\bar{\gamma}_{\rm M}\right)} {\bar{\gamma}_0+\bar{\gamma}_{\rm M} },
$$
which takes maximum value at
$$
\bar{\gamma}_{\rm M}=\alpha \left(1-\bar{\gamma}_0\right),
$$
with
$$
\alpha\equiv 1-\frac{2/3}{1+\sqrt{1-8\left(1-\bar{\gamma}_0\right)/9}} \in \left(\frac{1}{2},\frac{2}{3}\right).
$$
At the optimal $\bar{\gamma}_{\rm M}$, we have
\begin{align}
T^{-1}
& \sim \frac{a^2}{S_{\rm C}} \left(1-\bar{\gamma}_0\right)^{3}\frac{\alpha^2\left(1-\alpha\right)}{1-\left(1-\alpha\right)\left(1-\bar{\gamma}_0\right)}.
\nonumber \\
& \sim \frac{a^2}{S_{\rm C}} \left(1-\bar{\gamma}_0\right)^{3},
\nonumber
\end{align}
and the optimal time
$$
T^{\rm 2nd}_{\rm opt}
\sim \frac{S_{\rm C}}{a^2} \left(1-\frac{2{\gamma}_0S_{\rm C}}{a^2}\right)^{-3}.
$$
Above we have assumed $\gamma_{\rm M}$ could take any positive value but in the reality it is bounded by $\gamma^{\rm max}_{\rm M}$.
In the case
$$
\bar{\gamma}^{\rm max}_{\rm M}\equiv 2\gamma^{\rm max}_{\rm M}S_{\rm C}/a^2 \lesssim 1-\bar{\gamma}_0,
$$
the data acquisition time is optimal at $\gamma_{\rm M}=\gamma_{\rm M}^{\rm max}$, which is
$$
T^{\rm 2nd}_{\rm opt}
\sim\frac{1}{\gamma_{\rm M}^{\rm max}} \frac{1+\gamma_0/\gamma_{\rm M}^{\rm max}}{1-{2\gamma_0S_{\rm C}}/{a^2}}.
$$

\subsection{Optimal time for sensing based on fourth-order correlations}
\label{Append_sec_optimal_time_4th}

The optimal condition can be easily derived by solving the equation $\partial T/\partial \gamma_{\rm M}=0$.
The solution varies from
$\gamma_{\rm M}=a^2/\left(8S_{\rm C}\right)$ to $3a^2/\left(8S_{\rm C}\right)$ for
$\gamma_0$ varying from $0$ to $\infty$. Using the solution $\gamma_{\rm M}=a^2/\left(8S_{\rm C}\right)$  and neglecting the factors $\sim O(1)$ (such as the term $e^{8\gamma_{\rm M}S_{\rm C}/a^2}$),
we obtain the optimal data acqusition time as
$$
T_{\rm opt}^{\rm 4th}\sim \frac{8S_{\rm C}}{a^2}\left(1+\frac{8\gamma_0 S_{\rm C}}{a^2}\right)^2.
$$
Considering the physical constraint $\gamma_{\rm M}\le \gamma_{\rm M}^{\rm max}$, when  $\gamma_{\rm M}^{\rm max}\lesssim a^2/\left(8S_{\rm C}\right)$,
the data acquisition time is optimal at $\gamma_{\rm M}=\gamma_{\rm M}^{\rm max}$, being
$$
T_{\rm opt}^{\rm 4th}\sim \frac{1}{\gamma_{\rm M}^{\rm max}}\left(1+\frac{\gamma_0 }{\gamma_{\rm M}^{\rm max}}\right)^2.
$$
}

\end{appendix}

%


\end{document}